\documentclass[aps,preprint,showpacs,preprintnumbers,floatfix]{revtex4}

\usepackage{graphicx}% Include figure files
\usepackage{dcolumn}% Align table columns on decimal point

\begin{document}
\font\smallcap=cmcsc10

%\draft

\title
{Dynamic screening in $L_{2,3}$-shell transition metal x-ray absorption}

\author {A.\ L. Ankudinov}
\author {A.\ I. Nesvizhskii}
\altaffiliation[now at] {\ Institute for Systems Biology, Seattle WA 98103}
\author {J.\ J. Rehr}

\affiliation{Dept. of Physics, Box 351560,  University of
Washington, Seattle, Washington 98195}

\date{\today}
%\maketitle

\begin{abstract}
Calculations of $L$-shell x-ray absorption in transition metals
are shown to be sensitive to screening, both of the
x-ray field and the photoelectron-core hole interaction.
This screening is calculated using a generalization of the time dependent local
density approximation and a projection onto a local atomic basis.
The approach yields renormalized dipole-matrix elements which account
for the observed deviations of the $L_3/L_2$
branching ratio from the 2:1 value of independent electron theory.

\end{abstract}


\maketitle

%\section{Introduction}

Independent-electron theory is generally successful in describing
near edge x-ray absorption spectra (XAS) \cite{rehralbers}.
However, it fails dramatically at the $L_{2,3}$ edges in 
3d transition metals \cite{zaanen,waddington,thole88,swebert98}.
While the independent electron approximation predicts an $L_3/L_2$
transition intensity ``branching ratio" close to 2:1, the observed ratio
(Fig.~1) varies considerably with respect to atomic number $Z$, and is
closer to 1:1 for metals like Ti and V with nearly empty $d$-bands
\cite{leapman80,fink,jbarth}. This puzzling behavior is thought to
reflect many-body effects due to the Coulomb interaction, but
despite many studies, its variation has never been quantitatively
explained.  We now show, however, that the observed branching-ratio
depends crucially on dynamic screening of the x-ray field and
the photoelectron--core hole interaction, and can be calculated
using a simplified dynamic screening model and
a generalization of the time-dependent local density approximation (TDLDA).
Our approach makes use of a projection onto a local atomic basis
and a real-space multiple-scattering (RSMS) formalism \cite{arrc}.
This yields an efficient matrix formulation for extended systems, yet retains
the simplicity of the TDLDA, and gives results in
good agreement with experiment.

The $L_{2,3}$ XAS corresponds to transitions from the $2p_{1/2}$ and
$2p_{3/2}$ levels to continuum $s$ and $d$ states.
Several many body effects can be identified which contribute to
a non-constant $L_3/L_2$ intensity branching ratio:
i) Inelastic losses -- these can be represented
in terms of lifetime and self-energy effects in independent-electron
calculations.  The lifetimes are different for the $L_2$ and $L_3$ edges
due to the Coster-Kronig mechanism \cite{keski}, but this difference
only increases the branching ratios, e.g., to about 3:1.\
ii) Dynamic core polarization -- i.e., the creation of local
fields which screen the external x-ray field. This polarization effect
may be treated \cite{zs,swebert98} within the TDLDA by neglecting 
exchange terms, an approximation
often referred to as the random phase approximation (RPA).
This leads to a considerable reduction of the
branching ratio, but does not account well for its variation with $Z$.\
iii) Screening of the photoelectron-core hole interaction -- this
effect, which we find to be crucial, can be addressed in terms of
a frequency dependent exchange-correlation kernel $f_{xc}(\omega)$ 
in the TDLDA \cite{grosskohn}, or by the analogous, non-local
dynamically screened particle-hole interaction in the
Bethe-Salpeter equation (BSE) \cite{rohlfinglouie,soininen,strinati,hanke80}.
The importance of dynamic screening of the core-hole is surprising, since
it has been argued variously that corrections to the RPA are small
\cite{zs,swebert98}, or that an adiabatic kernel $f_{xc}^0$ is often
adequate \cite{zs,grosskohn}.
%Our results show that these simplifications cannot be taken for granted.

\begin{figure}
\includegraphics[width=8.2cm]{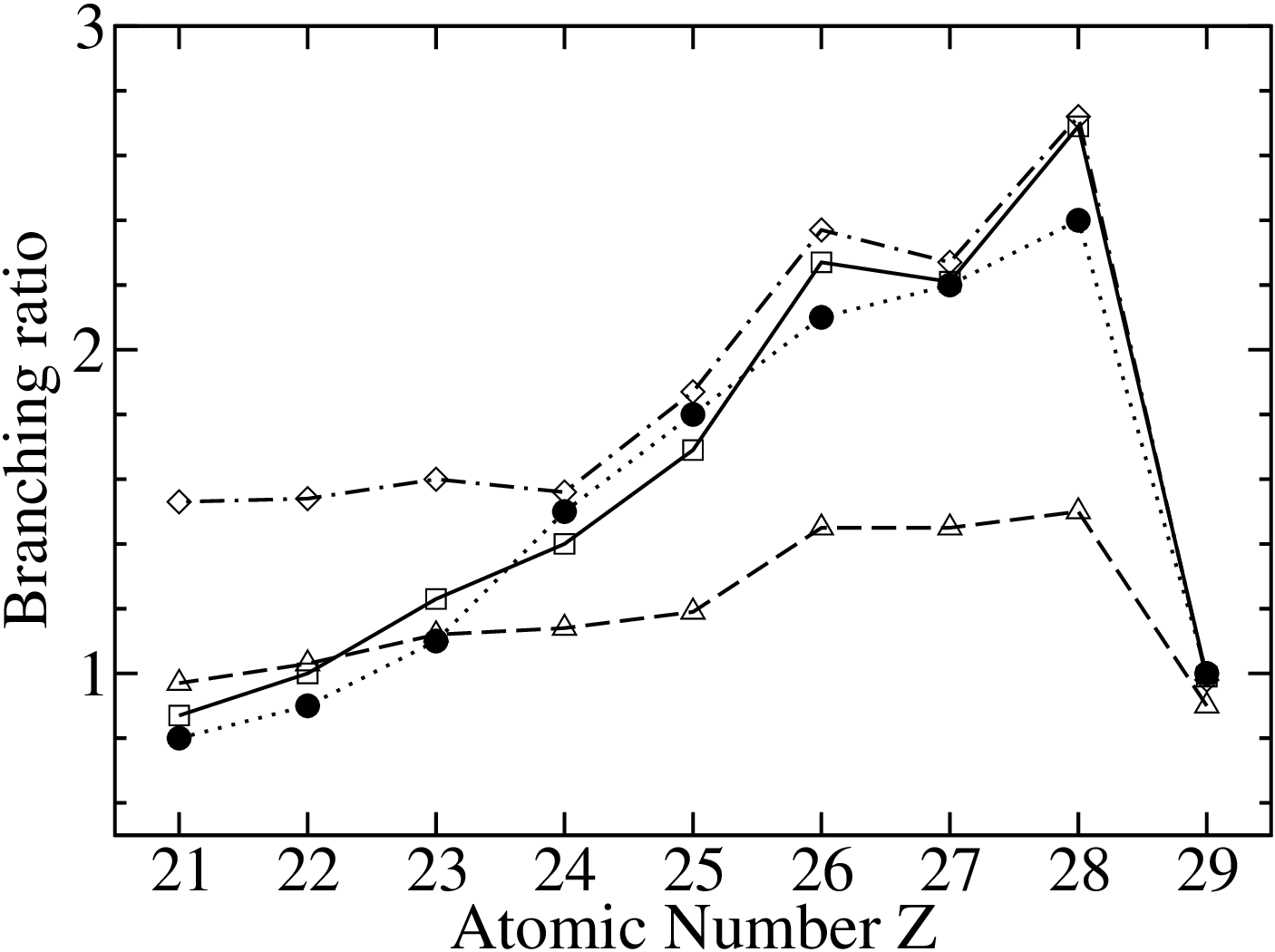}
\caption{\label{fig2}
 $L_3/L_2$ intensity branching ratio for the transition metal series
from experiment (solid circles), and as calculated with different
exchange correlation kernels: RPA (triangles),
adiabatic $f_{xc}^0$ (diamonds); and the dynamic model of this
work (see text) $\tilde f_{xc}(\tilde\omega)$ (squares).
}
\end{figure}

%\section{Local Basis TDLDA}

The TDLDA \cite{zs,swebert98,chelikowsky} provides an efficient
formalism for calculations of response functions, including corrections
to the independent electron approximation, since it avoids the
complications of non-locality in the time-dependent Hartree-Fock
(TDHF) \cite{crljen87}, BSE, or configuration interaction approaches.
 The TDLDA was originally introduced for atoms
\cite{zs}, but has since been extended to many other systems
\cite{chelikowsky} including band-structure formulations
for transition metals \cite{swebert98}. 
The TDLDA and TDHF equations are closely similar to the
BSE \cite{rohlfinglouie,soininen,strinati,hanke80}, which provides a
systematic, many-body framework based on the 2-particle Green's function
for treating core-hole screening.
The crucial difference between the TDLDA and the BSE lies in the
structure of the exchange-correlation kernel $f_{xc}(\omega)$; in addition
the single-particle states are replaced by quasiparticle states
which take the electron self-energy into account.  Our approach (see below)
is an extension of the TDLDA derived in part from the BSE.
Like the BSE, our approach also uses a quasiparticle approximation for
single-particle states, with inelastic losses approximated by a
Hedin-Lundqvist electron-gas self energy \cite{lundqvist69}. Such losses
are crucial in XAS and also point to the importance of dynamic screening.
Our method simplifies the screening calculations by using a local basis.
Local basis set methods have also been
used to advantage in various related calculations \cite{aryas95,rohlfing93}.

Within the TDLDA \cite{zs} or TDHF \cite{crljen87} approximations,
the photon cross-section (or XAS) $\sigma(\omega)$
can be expressed as an integral over the non-interacting response function
$\chi_0(\vec r,\vec r,'\omega )$ and the screened x-ray field
$\phi(\vec r,\omega).$
For notational simplicity, it is convenient to regard the continuous 
coordinates $\vec r$ and $\vec r\,'$ as vector or matrix indices,
which may be suppressed unless needed for clarity.
Then the XAS can be expressed compactly as 
\begin{equation}
\sigma(\omega)= -{4 \pi \omega \over  c } \phi^{*}(\omega)
               [ {\rm Im}\, \chi_0(\omega ) ]
               \phi(\omega),
 \label{muzs}
\end{equation}
where
\begin{equation}
   \chi_0(\vec r, \vec r,' \omega) =\sum_{ij}(f_i-f_j)
    {\psi_{i}^{*}(\vec r\,) \psi_{i}(\vec r\,')
    \psi_{j}^{*}(\vec r\,') \psi_{j}(\vec r\,) \over
    \omega + E_i - E_j + i 0^+} .
   \label{eq:chi0gf}
\end{equation}
Eq.~(1) is equivalent to an analogous expression with $\phi$ replaced
by the external x-ray field $\phi^{ext}$
and $\chi_0$ by the full response function $\chi$ \cite{zs}.
Here $f_i$ are Fermi occupation numbers (1 or 0), and the sums
run over all one-electron eigenstates $\psi_i(\vec r\,)$ of the ground
state Hamiltonian.  The field $\phi(\omega)$ consists of the external field
$\phi^{ext}\equiv\hat\epsilon\cdot\vec r$ (in the dipole approximation)
plus an induced local field, which in matrix form is given by
\begin{equation}
\label{eq:fieldscfK}
\phi(\omega)=\epsilon^{-1}(\omega) \phi^{ext}(\omega),\quad
\epsilon(\omega) = 1 - K(\omega) \chi^0(\omega).
\end{equation}
Here
$K(\vec r,\vec r,' \omega )$ denotes the particle-hole interaction (or
TDLDA kernel), which contains direct and exchange parts, i.e.,
\begin{equation}
   \label{eq:K}
   K(\vec r, \vec r,'\omega) = V(\vec r,\vec r\,') +
   f_{xc}(\vec r,\vec r,'\omega),
\end{equation}
and $V= 1/|\vec r - \vec r\,'|$ is the Coulomb interaction.

In this paper we consider several approximations for $f_{xc}(\omega)$,
which is generally a non-local, frequency dependent operator:
i) The RPA ($f_{xc}=0$)  -- to the extent exchange can be neglected, the
RPA is adequate \cite{swebert98}.\
ii) Adiabatic TDLDA ($f_{xc}(0)=f_{xc}^0$) -- this static limit
$f_{xc}^{0}(\vec r,\vec r\,') = \delta(\vec r  - \vec r\,') 
{\delta v_{xc}[\rho(\vec r\,)]/ \delta \rho }$, is dependent on the
local density and is obtained from the ground-state LDA
exchange-correlation potential $v_{xc}[\rho]$.\ 
iii) Dynamic TDLDA model --
 An LDA for the frequency dependence of $f_{xc}(\omega)$ 
was developed by Gross and Kohn \cite{grosskohn}.  At the large x-ray energies
of interest here, this $f_{xc}(\omega)$ is strongly suppressed, and yields
results close to the RPA \cite{swebert98}. However, such results are
clearly at odds with experiment for nearly empty $d$-bands (Fig.~1).\
iv) Dynamic TDLDA/BSE model -- Our aim here is to improve on i), ii) and iii)
for $L_{2,3}$ XAS, based partly on the BSE
\cite{rohlfinglouie,soininen,strinati,hanke80}.
In the BSE, the matrix elements
$\langle  vc \vert f_{xc}(\omega) \vert v'c' \rangle$
depend on the dynamically screened Coulomb interaction
$W(\omega)=\tilde\epsilon^{-1}(\omega)V$, through an
effective inverse dielectric matrix $\tilde\epsilon^{-1}(\omega)$
\cite{strinati}.
%reduces to the unscreened limit $\delta(\vec r -\vec r\,')$ at high energy.
However the actual dependence on $\omega\approx E_c-E_v$
is matrix element dependent, and depends on the effective dielectric
response at the energy-transfer frequency, i.e.,
$\tilde\omega=\omega+E_{c'}-E_{v}\approx E_{v'}-E_{v}$.  This behavior
can be seen explicitly in plasmon-pole models \cite{rohlfinglouie}.  
For $L_{2,3}$ XAS, the most important occupied states $v,v'$ are 
the $2p_{1/2}$ and $2p_{3/2}$ levels, which are split
by a moderate spin-orbit interaction $\Delta_{so}$, ranging from 5 eV for Sc
to 20 eV for Cu.  The matrix elements with zero energy transfer correspond
to static screening; thus it is reasonable to set
$f_{xc}(\omega) = f_{xc}^0$ for $v=v'$.  We
also tried the unscreened, non-local TDHF exchange operator
for $v=v'$, which corresponds to an unscreened
core-hole potential, but found it to be much too strong.
For the off-diagonal elements ($v\neq v'$), however, we found
that the unscreened exchange operator (i.e., the high frequency
limit $W=V$) has only a small effect.
This suggests that the effects of dynamic screening on off-diagonal terms
at moderately high frequency $\tilde\omega=\Delta_{so}$ are also small and
can be neglected.  This behavior is in contrast to the case for optical
absorption, where the adiabatic limit ($\tilde\omega=0$) is a good
approximation and can be used for all matrix elements \cite{rohlfinglouie}.
Thus remarkably, we find that elaborate calculations of dynamical
screening can be avoided for $L_{2,3}$ XAS by using a simplified dynamic model 
$f_{xc}(\omega)\rightarrow \tilde f_{xc}(\tilde\omega),\
 \tilde\omega=E_v-E_{v'}$, i.e.,
$\tilde f_{xc}(\tilde\omega)=f_{xc}^0$ for $\tilde\omega=0$ and 
$\tilde f_{xc}(\tilde\omega)=0,\ (\tilde\omega=\Delta_{so})$
This defines our dynamic TDLDA/BSE model, which 
leads to reasonable agreement with experiment (Fig.~1). 

Next we briefly outline our calculations, which make use
of the RSMS formalism (i.e., the real-space analog of the Koringa-Kohn
Rostoker (KKR) band structure method) of our self-consistent,
all-electron FEFF8 code \cite{arrc}. To begin we rewrite Eq.~(1) as
\begin{equation}
    \sigma(\omega) = {4 \pi e^2 \omega \over  c } \sum_{v,LL'} 
    \tilde M_{vL}(\omega)  \rho_{L, L'} (E)
    \tilde M_{vL'}(\omega),
    \label{muequiv}
\end{equation}
where $E=\omega+E_v-E_F$ is the photoelectron energy.
The screening of both the x-ray field and the
photoelectron-core hole interaction are included implicitly in
the renormalized dipole matrix elements \cite{crljen87},
$\tilde M_{vL}(\omega ) =\langle R_L |\phi|v\rangle$, where
$L=(\kappa,m)$ denotes a relativistic angular momentum basis.
 The quantities $\rho_{L, L'} (E)$ are matrix elements of 
the unoccupied one-electron density matrix, 
\begin{eqnarray}
\rho(\vec r,\vec r\,',E) &\equiv&
\sum_c \psi_{c}(\vec r\,) \psi_{c}^{*}(\vec r\,')
\delta(E-E_c), \nonumber \\
&=& \sum_{L,L'} R_L(\vec r\,)R_{L'}(\vec r\,') \rho_{L, L'}(E),
\nonumber \\
\rho_{L,L'}(E) &=& \delta_{L, L'} + \chi_{L, L'}(E).
\end{eqnarray}
Here $R_L(\vec r\,,E)$ are normalized scattering states calculated with
the absorbing atom potential, and $\chi_{L,L'}(E)$ contains the fine
structure in the XAS due to scattering by the environment \cite{arrc}.
Note that by replacing $\phi$ with $\phi^{ext}$ in Eq.~(5),
the screened dipole matrix
elements $\tilde M_{vL}$ become bare dipole matrix elements
$M_{vL}=\langle R_L |\hat\epsilon\cdot\vec r\,|i\rangle$, and one recovers the
independent electron formula, equivalent to Fermi's Golden Rule.
Since the strength of the XAS is a measure of the screening response,
the independent-electron approximation should become
increasingly valid away from the edge region.

The second key approximation in our approach is the use of a local basis
for calculations of $\chi_0$ and $\tilde M_{vL}$.
This is done starting from an expression in terms of a
Kramers-Kronig (KK) transform over the density matrix, 
\begin{eqnarray}
   \label{eq:chi0final}
   \chi_0 (\vec r, \vec r\,', \omega) &=& 
   \sum_{v} \psi_{v}^{*}(\vec r\,) \psi_{v}(\vec r\,')
   \int_{E_F}^{\infty} \frac{d E}{\pi}\,
   \rho(\vec r, \vec r,'E) \\ \nonumber 
   &\times& \left[ {1\over\omega - E + E_v  + i\delta}
                  +{1\over\omega + E - E_v  + i\delta}\right].
\end{eqnarray}
Once $\chi_0$ is known, Eq.~(\ref{eq:fieldscfK}) could be solved
iteratively in real space to obtain $\phi(\vec r\,)$ \cite{zs}.
However, this procedure is computationally
expensive for extended systems, since it involves KK transforms for many
$(\vec r, \vec r,' \omega)$ points.  To simplify this calculations,
we make the reasonable assumption \cite{zs} that the induced charge
$\rho^{ind} = \chi_0(\omega) \phi$ that screens the x-ray field,
is local and arises largely from a few significant
orbitals on the absorbing atom. This is convenient, since our formulation 
only needs the screened field $\phi(\vec r,\omega)$ at
short distances to calculate the deep-core transition matrix
$\tilde M_{vL}(\omega)$.
Thus to approximate $\phi$, we introduce the atomic projection operator
${P} = \sum_n |\psi_n \rangle \langle \psi_n|$, which projects a given
function onto a local basis set of atomic-like orbitals on the central atom.
Then the density matrix can be approximated by its local contribution
$\rho \approx {\rho}^{loc} = {P} \rho {P}$. These approximations can
be systematically improved by including a more complete set. Thus
\begin{eqnarray}
   \label{eq:chimatrix}
\chi_{0}^{loc}(\vec r, \vec r,'\omega) &=& 
   \sum_{vnn'} \psi_v^*(\vec r\,) \psi_n^*(\vec r\,') 
   {\chi}_{vn,vn'}^{loc}(\omega) \\
  &\times &\psi_{v}(\vec r\,') \psi_{n'}(\vec r\,), \nonumber \\
{\chi}_{vn,vn'}^{loc}(\omega) &=&  - \frac{k}{\pi}
   \sum_{L,L'} \int_{E_F}^{\infty} d E \, 
   \frac{ \langle n | R_{L} \rangle  \rho_{L,L'}
    \langle  R_{L'}  | n' \rangle }
    {\omega -E +E_v + i \delta},\,\,  \nonumber
\end{eqnarray}
where $k=\sqrt{2(\omega+E_v)}$.  Note that the localized part of $\chi_0$
does not require a KK transform at each point,
since the localized part of the photoelectron wave function can 
be separated into energy and position dependent parts.  Moreover, the
overlap matrices $\langle n | R_{L}\rangle$ decay rapidly with energy, so the
KK transform converges well.  This approximation then leads to
a fast matrix formulation for $\tilde M_{vL}$.
 From Eq.~(\ref{eq:fieldscfK}), we obtain
(summation over repeated indices being implicit)
\begin{eqnarray}
   \label{eq:fieldscfKloc}
   \phi(\vec r, \omega) &\approx& \phi^{ext} (\vec r , \omega) +  
   \sum_{v'n'n''} \int d \vec r\,' \,  K(\vec r, \vec r,' \omega )
   \nonumber \\
   &\times& \psi_{v'}^{*}(\vec r\,')
   \psi_{n'}(\vec r\,') \: {\chi}_{v'n',v'n''}^0(\omega)  
   \tilde M_{v'n''} \:, 
\end{eqnarray}
where $\tilde M_{vn} = \langle \psi_{n} | \phi | v \rangle $
is calculated by  integrating Eq.~(\ref{eq:fieldscfKloc})
over the core and basis set functions,
\begin{eqnarray}
   \label{eq:Klocmat}
   \tilde M_{vn}(\omega) &=& M_{vn} +
   K_{vn,v'n'}\,\tilde{\chi}^{loc}_{v'n',v'n''}(\omega)\,
   \tilde M_{v'n''}(\omega), \nonumber \\
M_{vn}&=&\langle n | \phi^{ext} | v \rangle,\quad
K_{vn,v'n'} = \langle vn| K |v'n'\rangle.
\end{eqnarray}
These equations can readily be solved by matrix inversion.
Finally on integrating Eq.~(\ref{eq:fieldscfKloc}) over the core- and
final state-wave functions, we get 
\begin{eqnarray}
  \label{eq:mli}
  \tilde M_{vL}(\omega) &=&  M_{vL}(\omega) +
   K_{vL, v'n'}\chi^{loc}_{vn',v'n''}\: \tilde M_{v'n''},  \nonumber \\
   K_{vL,v'n'} &=& \langle v R_L | % R_L(\omega-E_v)|
     K(\omega)| v'n'\rangle,  
\end{eqnarray}
where $R_L$ denotes the scattering-state $R_L(\omega-E_v)$.
The matrix form in Eq.~(\ref{eq:mli}) is very efficient,
and has been implemented using an extension of our FEFF8 code.  This extension
is straightforward, since only the dipole matrix elements need be modified
to incorporate screening.  Due to the local form of
$\tilde f_{xc}(\tilde\omega)$,
the contributions to $K_{vL,v'n'}(\omega )$ satisfy the same selection rules
and can be calculated using standard formulas for Coulomb interaction
matrix elements \cite{grant}.

\begin{figure}
\includegraphics[width=8.2cm]{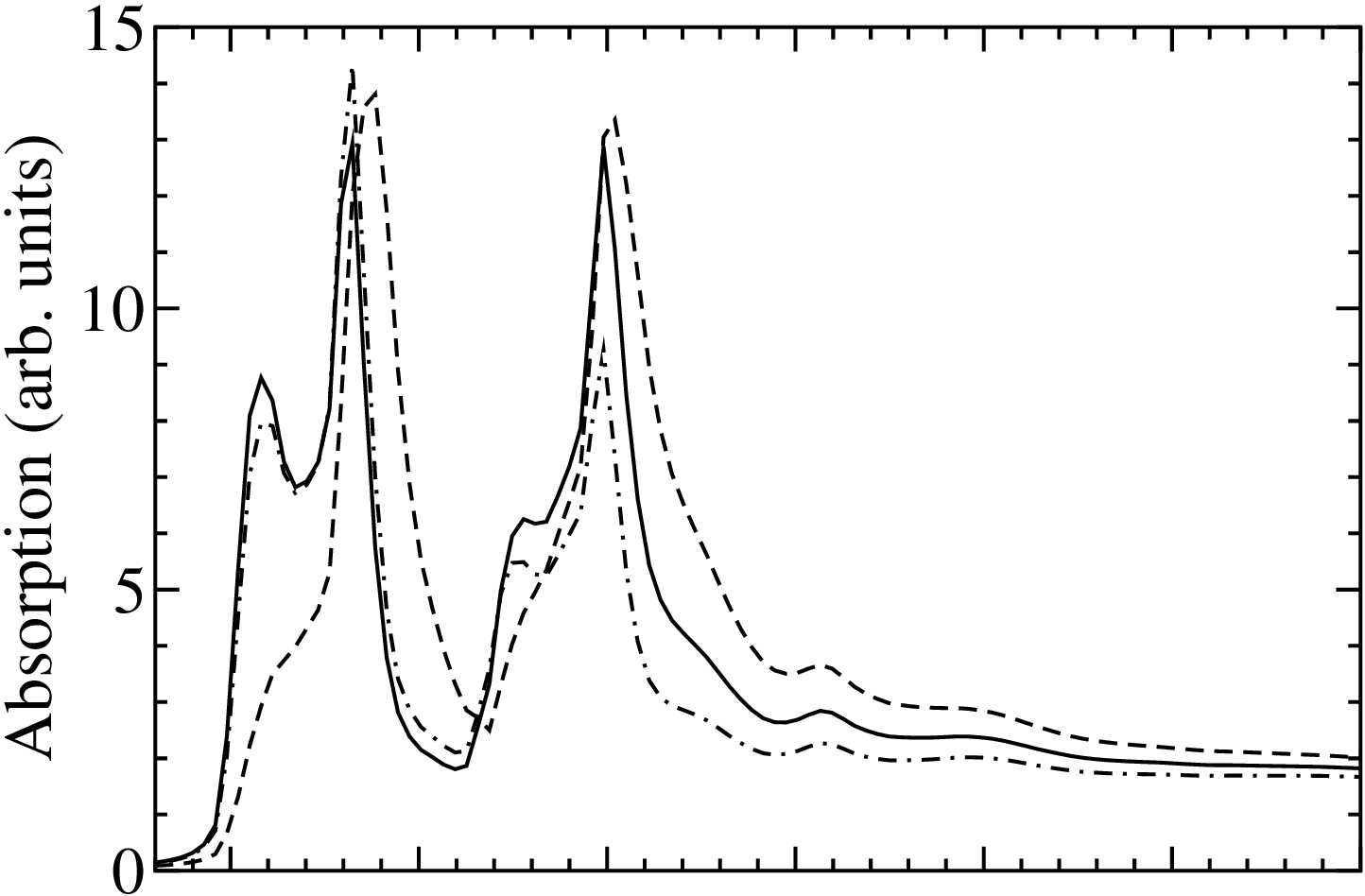}
\vskip 0.85truecm
\includegraphics[width=8.2cm]{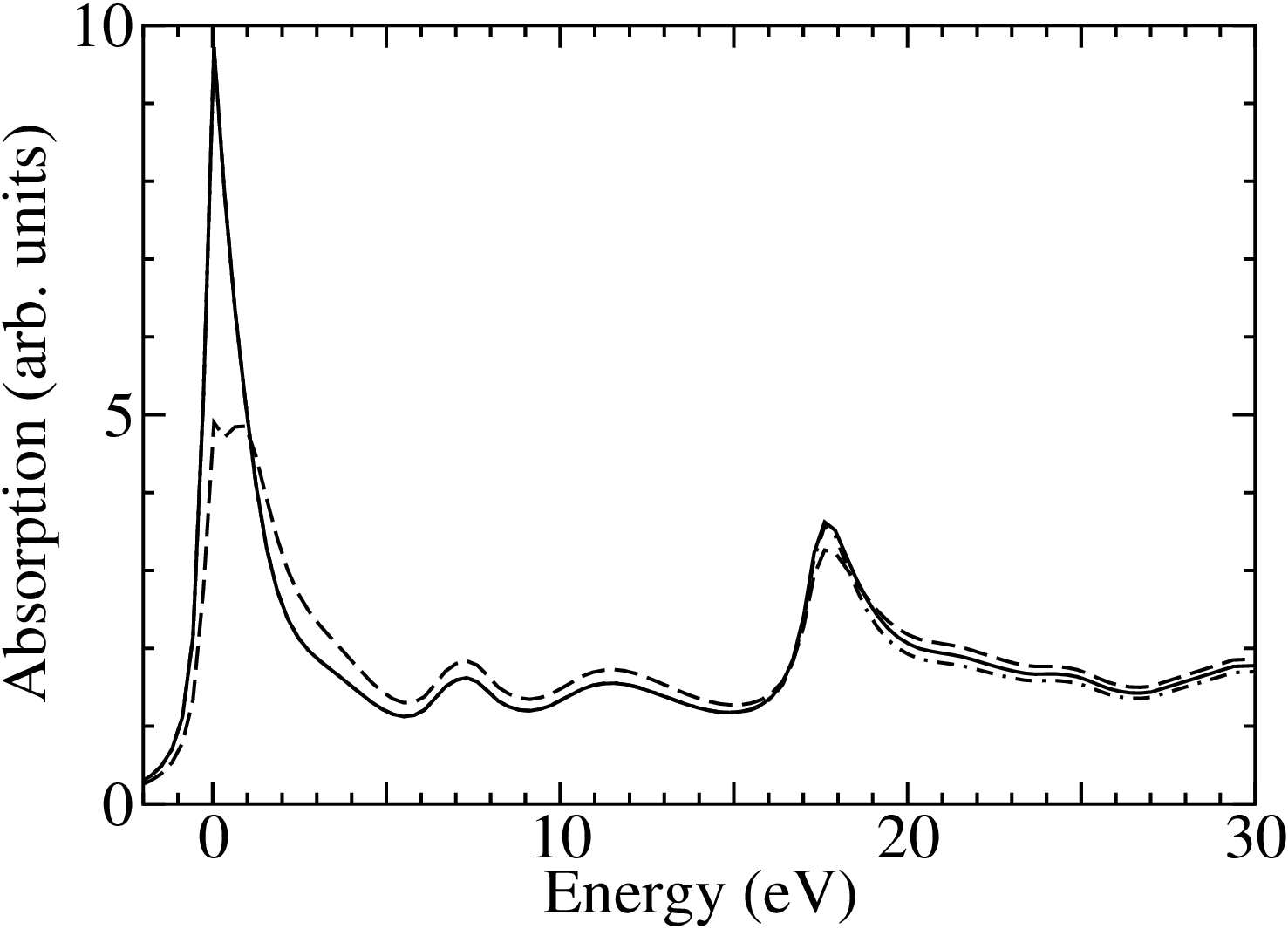}
\caption{\label{fig1ti} L$_{3,2}$ edge XAS vs energy with respect
to the Fermi level, for a) Ti (upper figure)
and b) Ni (lower), as calculated with different screening models:
RPA  (solid), static $f{xc}^0$ (dots), and the dynamic model
of this work $\tilde f_{xc}(\omega)$ (dashes).
}
\end{figure}

%\section{Comparison with Experiment}
Typical results near the  beginning and end of the
transition metal series are presented for Ti  and Ni in Fig.~2.
The dramatic  differences reflect differences in the response
between nearly empty and nearly filled $d$-bands, and are strongly
dependent on the form of $f_{xc}(\omega)$.
For all calculations we used theoretical atomic
core-hole life-times \cite{keski}.
We did not add additional broadening to correct for experimental resolution,
though this would give slightly better agreement with experiment (Fig.~1).
Our results for the RPA agree well with those of Ref.~[\onlinecite{swebert98}],
which validates our local screening approximation.
Note that the RPA is only a good approximation for nearly empty $d$-bands,
while the adiabatic $f^0_{xc}$ is appropriate only for nearly filled ones.
However, our dynamic model $\tilde f_{xc}(\tilde\omega)$ is clearly
satisfactory for the entire series.

Since screening redistributes the oscillator strength between the $L_2$ and
$L_3$ edges, the importance of these many-body corrections appears to cast
doubt on the accuracy of results obtained from the XAS ``sum-rules"
\cite{swebert98}.  These sum-rules allow one to determine various spin- and
orbital-moments from linear combinations of the $L_2$ and $L_3$
XAS \cite{nesvisr}.  However,
one can now correct these procedures for local-field effects with our approach,
e.g., by substituting the screened XAS cross-section, in place of the
one-electron result in the analysis.

In summary, we have found that dynamic screening of the photoelectron-core
hole interaction, which gives rise to a frequency dependent
exchange-correlation operator $f_{xc}(\omega)$, is crucial in calculations
of transition metal $L_{2,3}$ spectra. However, we have found a dynamic model
based on the TDLDA and BSE, which accounts well for the frequency and
matrix element dependence, by neglecting off diagonal, high frequency
screening terms.  With this model, we have developed  an
efficient approach for including screening in deep-core XAS,
based on calculations of screened dipole matrix elements.
Our approach goes beyond the conventional TDLDA, and is similar in some
respects to a screened TDHF approximation \cite{hanke80}. Moreover, the
approach yields good agreement with experiment for the $L_{2,3}$ XAS of 3d
transition metals, without the complexity of full dynamic-screening
calculations.

%Acknowledgments:

\begin{acknowledgments}
We thank J. Chelikowsky, H. Ebert, W. Ku, Z. Levine, 
S. Pantelides, G. Sawatzky, E. Shirley and especially G. Bertsch and
A. Soininen for helpful comments. 
This work was supported by DOE grants DE-FG03-97ER45623 and
DE-FG03-98ER45718, and was
facilitated by the DOE Computational Materials Science Network.
\end{acknowledgments}

\end{document}